\documentclass[10pt,conference]{IEEEtran}
\IEEEoverridecommandlockouts

\usepackage{cite}
\usepackage{amsmath,amssymb,amsfonts}
\usepackage{algorithmic}
\usepackage{graphicx}
\usepackage{textcomp}
\usepackage{xcolor}
\usepackage{hyperref}
\usepackage{makecell}
\usepackage{url}
\usepackage{booktabs}
\usepackage[detect-none]{siunitx}
\def\BibTeX{{\rm B\kern-.05em{\sc i\kern-.025em b}\kern-.08em
    T\kern-.1667em\lower.7ex\hbox{E}\kern-.125emX}}

\newcommand{\fix}[1]{\textcolor{black}{#1}}

\begin{document}

\title{\textsc{AILinkPreviewer}: Enhancing Code Reviews with LLM-Powered Link Previews}

\author{\IEEEauthorblockN{Panya Trakoolgerntong\IEEEauthorrefmark{1}, Tao Xiao\IEEEauthorrefmark{2}, Masanari Kondo\IEEEauthorrefmark{2}, Chaiyong Ragkhitwetsagul\IEEEauthorrefmark{1},\\ Morakot Choetkiertikul\IEEEauthorrefmark{1}, Pattaraporn Sangaroonsilp\IEEEauthorrefmark{1}, Yasutaka Kamei\IEEEauthorrefmark{2}}
\IEEEauthorblockA{\IEEEauthorrefmark{1}Faculty of Information and Communication Technology, Mahidol University, Nakhon Pathom, Thailand\\
\IEEEauthorrefmark{2}Kyushu University, Fukuoka, Japan\\Email: panya.tra@student.mahidol.ac.th, \{chaiyong.rag, morakot.cho, pattaraporn.san\}@mahidol.ac.th,\\ \{xiao, kondo, kamei\}@ait.kyushu-u.ac.jp}
}

\maketitle

\begin{abstract}
Code review is a key practice in software engineering, where developers evaluate code changes to ensure quality and maintainability. Links to issues and external resources are often included in Pull Requests (PRs) to provide additional context, yet they are typically discarded in automated tasks such as PR summarization and code review comment generation. This limits the richness of information available to reviewers and increases cognitive load by forcing context-switching. To address this gap, we present \textsc{AILinkPreviewer}, a tool that leverages Large Language Models (LLMs) to generate previews of links in PRs using PR metadata, including titles, descriptions, comments, and link body content. We analyzed 50 engineered GitHub repositories and compared three approaches: Contextual LLM summaries, Non-Contextual LLM summaries, and Metadata-based previews. The results in metrics such as BLEU, BERTScore, and compression ratio show that contextual summaries consistently outperform other methods. However, in a user study with seven participants, most preferred non-contextual summaries, suggesting a trade-off between metric performance and perceived usability. These findings demonstrate the potential of LLM-powered link previews to enhance code review efficiency and to provide richer context for developers and automation in software engineering.

The video demo is available at~\url{https://www.youtube.com/watch?v=h2qH4RtrB3E}, and the tool and its source code can be found at \url{https://github.com/c4rtune/AILinkPreviewer}.
\end{abstract}


\section{Introduction}
\label{sec:introduction}
Code review is one of the key collaborative activities in software development \cite{10.1145/3274404}.
It enables developers to evaluate code changes for correctness and maintainability before integration. However, the process is often challenging due to time pressure, communication barriers, and the inherent complexity of code comprehension \cite{9793977, 8453136}. Prior work emphasized that reviewing code is not only a technical task but also a complex cognitive process, requiring developers to juggle multiple information sources while maintaining focus \cite{10.1145/3274404, 8094433, wang2021automatic, wang2023exploration}.

One key source of cognitive load arises when reviewers must navigate away from the code review environment to consult external resources through hyperlinks. This context switching introduces an extraneous cognitive load, as developers expend mental resources to manage task-switching mechanics rather than to understand the code itself~\cite{TheTrueC86:online}. Although links often provide crucial information needed during review, such as documentation or external resources \cite{article}. However, reviewers are forced to open and interpret these links manually, which can fragment attention and prolong review time. 

\begin{figure}[tb]
    \centering
\includegraphics[width=\columnwidth]{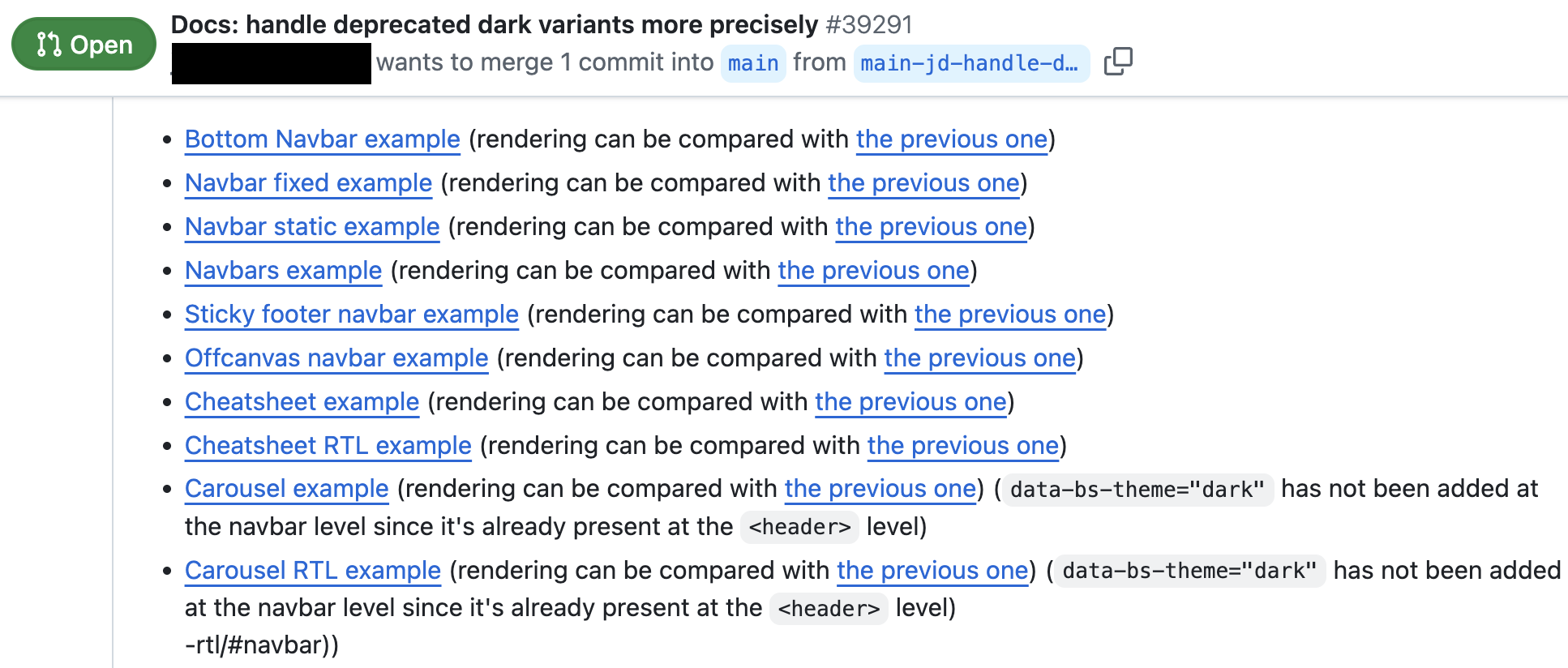}
    \caption{A PR containing numerous links to external resources}\label{fig:Many_link_pull_request}
\end{figure}

Figure~\ref{fig:Many_link_pull_request} shows an example that motivates our study. The PR \#39291 of the bootstrap project,\footnote{\url{https://github.com/twbs/bootstrap/pull/39291}}, which updates Bootstrap's documentation to clarify the handling of deprecated \texttt{.dark} variant classes in components like dropdowns and navbars, ensuring consistent guidance and pointing users toward the newer color-mode approach. It contains 59 links, 4 of which point to an internal link, while the remaining 55 links point to external resources. Thus, reviewers need to go through all the links and understand them to get the complete context for this PR. This is a tedious and cumbersome task.

Moreover, despite their importance, links are routinely discarded in automated software engineering tasks such as PR summarization and code review comment generation \cite{sakib2024automaticpullrequestdescription, shi2022are}. Removing these links strips away valuable contextual information that could enrich both automated tools and human review. Incorporating knowledge from links has the potential to mitigate cognitive load, improve review efficiency, and support more informed decision-making.

In this work, we propose \textsc{AILinkPreviewer}, an LLM-powered approach to generate link previews that integrate contextual information from PR metadata. By bridging the gap between hyperlinks and their surrounding context, our tool reduces the need for developers to leave the review environment, thereby lowering extraneous cognitive load and improving reviewer efficiency. This direct access to contextualized link information can lead to faster and more focused reviews during collaborative development. Beyond practitioner benefits, \textsc{AILinkPreviewer} also enriches automated software engineering research by retaining links that are typically discarded in summarization and comment generation tasks. Incorporating knowledge from links enables more accurate and context-aware automation in software engineering.

\section{Compared Methods}
\label{sec:cm}

To investigate the impact of LLM on link preview generation in code reviews, we designed three comparative methods. Each method produces summaries of hyperlinks appearing in PRs but differs in how contextual signals are incorporated. 

\textbf{Contextual LLM Summaries.} This method prompts an LLM to generate a summary of a hyperlink while leveraging contextual information surrounding the PR. The context includes the PR title, description, repository name, repository description, and the body content of the link itself. By grounding the summary in these additional metadata, the generated preview is expected to better reflect the information needs of reviewers. The specific context provided depends on the location of the link. If the link appears in the description, the description body is included; if it is embedded in a comment, the corresponding comment body is included; and if it occurs within a review comment, the review comment body is incorporated. This adaptive contextualization ensures that the generated summaries remain sensitive to the conversational or structural role of the hyperlink within the review process.

\textbf{Non-Contextual LLM Summaries.} The second method also uses the LLM, but without providing any contextual signals from the PR. The model receives only the body content of the link, and the generated summary reflects the information present in the link in isolation. This method provides a baseline for assessing the contribution of contextual information.

\textbf{Metadata-Based Snippet Summaries.}
The third method does not rely on an LLM. Instead, it combines the hyperlink's title and description metadata to create a summary, similar to the snippets commonly shown in search engines, e.g., Google \cite{google_featured_snippets, google_search_help_snippets, webis_abstractive_snippet}. This lightweight baseline shows the utility of metadata in providing reviewers with at-a-glance information.

These three approaches allow us to systematically compare the role of contextual information in link preview generation. The Contextual LLM Summaries capture the richest set of signals, the Non-Contextual LLM Summaries isolate the contribution of the link itself, and the metadata-based snippets represent a simple heuristic baseline.

\section{Tool Architecture}

\begin{figure}[t]
    \centering
\includegraphics[width=\columnwidth]{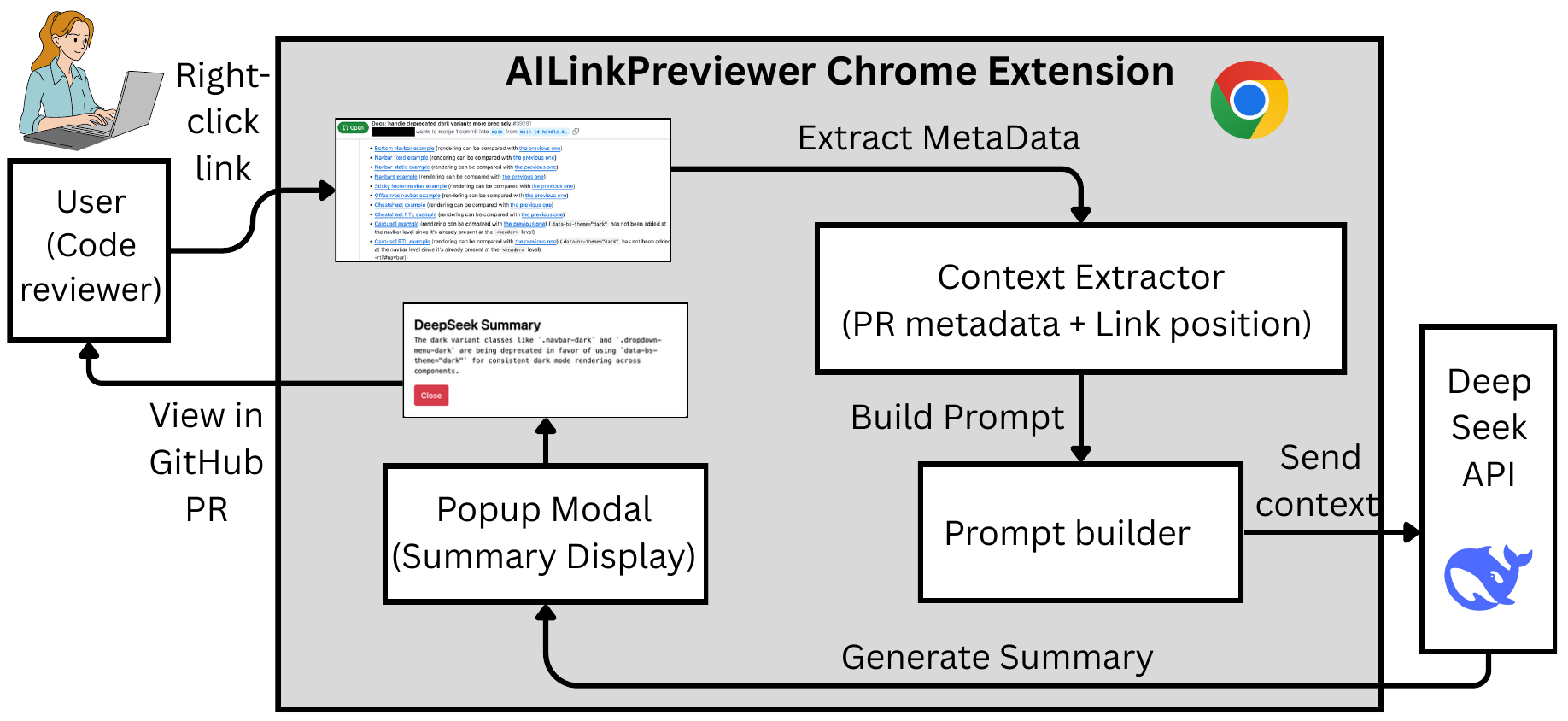}
    \caption{Architecture for \textsc{AILinkPreviewer} extension}
    \label{fig:extension_architecture}
\end{figure}

\begin{figure}[t]
    \centering
\includegraphics[width=0.7\columnwidth]{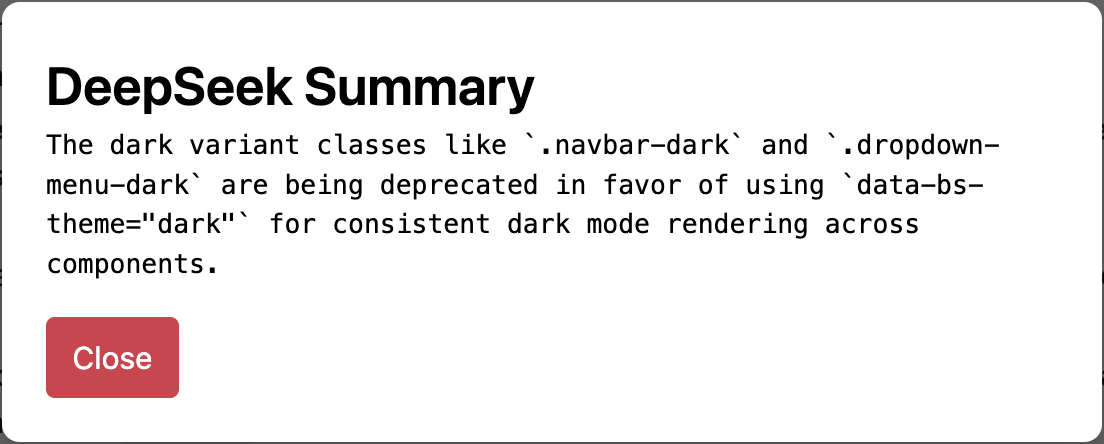}
    \caption{Example of \textsc{AILinkPreviewer} popup modal}
    \label{fig:pop_up_example}
\end{figure}

The system architecture of 
\textsc{AILinkPreviewer} is depicted in Figure~\ref{fig:extension_architecture}. The tool is designed to integrate seamlessly into the GitHub PR interface. The process begins with the user configuring the extension by providing their DeepSeek API key through the extension widget. Once the API key has been set, the extension becomes active within the browser environment. During a code review session, the user navigates to a GitHub PR and selects a hyperlink appearing in the PR content. By right-clicking on the link, a context menu is displayed, offering the option ``AILinkPreviewer: Summarize Link.'' When this option is chosen, the extension extracts relevant contextual information, including the PR title, PR description, repository title, repository description, and the metadata of the linked page. Importantly, the context provided to the summarization model is also influenced by the position of the hyperlink, as explained in Section~\ref{sec:cm}. This combined information is then passed to a DeepSeek-powered prompt, which generates a summary tailored to the code review context. Finally, the extension presents the generated summary in a pop-up modal as shown in Figure~\ref{fig:pop_up_example} overlaying the GitHub interface. This allows reviewers to access contextualized previews of linked resources without leaving the PR page, thereby reducing context switching and cognitive load.

We provide an open-source tool for easy deployment. Users only need to extract and load it into Google Chrome as an unpacked extension via Developer Mode.

\section{Dataset and Evaluation Setup}
\subsection{Dataset}
The initial dataset is derived from the curated list of GitHub projects provided by Dabic et al.~\cite{Dabic:msr2021data}. To select active and filter out toy projects, we applied the following filters: (i) at least 100 commits, (ii) at least one issue, (iii) at least three contributors, (iv) at least 100 PRs, (v) at least one release, and (vi) a last commit no earlier than May 27, 2024. In addition, we excluded forks and sorted the remaining candidates in descending order by their stars.

We manually reviewed the projects to ensure compliance with the definition of an engineered software project, namely a project that leverages sound software engineering practices in each of its dimensions, such as documentation, testing, and project management~\cite{10.1007/s10664-017-9512-6}. This process resulted in the top 50 highly active and mature repositories, which form the basis for our evaluation. To ensure the quality and relevance of the evaluation dataset, we targeted only hyperlinks explicitly written in Markdown format (e.g., \texttt{[label](url)}). Furthermore, we applied a length filter, retaining only links whose labels contained eight or more words. This criterion helped exclude trivial or overly terse labels (e.g., ``click here'') and ensured that the retained links carried sufficient descriptive content for meaningful summarization and comparison. \fix{We conducted a manual inspection of 365 samples (with a confidence level of 95\% and a confidence interval of 5), focusing on the choice of eight words. The findings indicate that employing eight words results in only 15.07\% false positives (labels that are trivial).}

\subsection{Evaluation Setup and Results}

\textbf{Model Accuracy.} To assess the quality of generated previews, we used a combination of automatic and human-centered evaluation metrics. We used the label of the link as reference text for the summarization of the link by developers.
For textual similarity and overlap, we used BLEU, ROUGE, and METEOR, which have been widely adopted in software engineering tasks such as commit message generation and PR summarization~\cite{liu2018neural,ahmad2020transformer}. To capture semantic alignment, we applied BERTScore and sentence similarity measures, both of which have shown effectiveness in evaluating natural language artifacts in code-related tasks~\cite{zhang2020bertscore,wang2021document}. Relevance to the body content of the link was measured to evaluate how well summaries reflected the information needs of code reviewers, following prior work on PR summarization and review support~\cite{shi2021automatic,liu2020automatic}. In addition, we assessed readability using the Flesch Reading Ease score~\cite{mallinson2017sentence} and conciseness using the compression ratio. Together, these metrics provide a multi-dimensional view of summary quality, balancing lexical overlap, semantic content, readability, and efficiency.

\begin{table}[tb]
\centering
\caption{Evaluation Results.}
\label{tab:auto_results}
\begin{tabular}{lrrr}
\toprule
Metric & CLS & NCLS & MBS \\
\midrule
BLEU&\textbf{4.52}&3.02&1.20\\
METEOR&\textbf{19.06}&15.85&12.50\\
ROUGE 1&\textbf{19.36}&16.69&12.50\\
ROUGE 2&\textbf{6.96}&4.78&2.69\\
Sentence Similarity&\textbf{14.63}&11.81&10.97\\
Flesch Reading Ease&32.28&\textbf{30.34}&39.57\\
BERT precision&\textbf{48.73}&47.49&39.83\\
BERT Recall&\textbf{55.47}&53.54&49.85\\
BERT F1 score&\textbf{51.67}&50.10&43.97\\
Compression ratio&\textbf{1.61}&1.67&2.04\\
Text relevance&\textbf{83.81}&82.03&83.34\\
\midrule
Preference from user study & 9 & \textbf{21} & 5\\
\bottomrule
\multicolumn{4}{l}{\scriptsize{Note: CLS = Contextual LLM summaries, NCLS = Non-Contextual}}\\
\multicolumn{4}{l}{\scriptsize{LLM summaries, MBS = Metadata-based Snippets}} \\
\end{tabular}
\end{table}%

\fix{Our evaluation uses several libraries and metrics from natural language processing research. NLTK was used for tokenization, BLEU, and METEOR calculations. ROUGE scores (ROUGE-1 and ROUGE-2 F1) were computed using the \texttt{rouge\_score} package, while semantic similarity was estimated by vectorization of TF-IDF and cosine similarity. Readability was measured using the Flesch Reading Ease formula provided by the \texttt{textstat} library. The compression ratio was computed as the length of the summary divided by the length of the body content. For semantic embedding-based evaluation, we employed BERTScore, implemented with Hugging Face's transformers library. We used the pretrained \texttt{bert-base-uncased} model to obtain token embeddings and computed precision, recall, and F1 scores based on cosine similarity between summary and reference embeddings. The pretrained model is also used to calculate the cosine similarity of texts between the summary and the body content of links.}

\textbf{Results in Model Accuracy.}
The final metrics reported in Table~\ref{tab:auto_results} demonstrate the comparative effectiveness of contextual information in improving summary quality. The table presents the average accuracy across 50 projects. The results indicate that the Contextual LLM Summary approach outperforms both the Non-Contextual LLM Summary and the Metadata-based snippet on nearly all measures. In terms of lexical overlap, Contextual LLM  Summary achieves 4.52 BLEU and 19.36 ROUGE-1, exceeding the Non-Contextual LLM Summary by 1.50 (BLEU) and 2.67 (ROUGE-1) points, respectively, and substantially surpassing the Metadata-based snippets. For semantic similarity, Contextual LLM Summary reaches 14.63 in sentence similarity and a BERT F1-score of 51.67, outperforming the Non-Contextual LLM Summary by 2.82 and 1.57 points, respectively. It also achieves the highest relevance score (83.81), demonstrating its ability to capture the contextual information most useful to reviewers. While the Metadata-based snippets show slightly higher readability (39.57 Flesch Reading Ease), this comes at the cost of lower accuracy and informativeness. These results confirm that incorporating contextual Metadata enables the Contextual LLM Summary to generate more accurate, relevant, and semantically rich previews than competing methods.

\textbf{Tool Usability.}  We invited 7 participants from our research group to test our tools. All participants have submitted their results from using the tools, but only 5 of them have submitted the survey afterward. Most of the participants have more than four years of programming experience \fix{and GitHub}. %

We provided an example of how to operate the tools in a README.md file, as well as a short video to demonstrate each step to use the tool. However, we did not provide a list of existing PRs for them. \fix{To mimic the review process, they} are free to test our tools in any PR. We rationalize based on how a real user would use the tool by selecting repositories that would be useful to them. Then, they can choose their preferred preview from the three approaches for each PR.

Finally, we asked them to provide a score rating on two aspects: (1) Ease-of-use: the tool is easy to use; (2) Usefulness:
The tool is useful and can help you with your task. Each score is based on the Likert scale ranging from \numrange{1}{5}.

\textbf{Results in Tool Usability.} The results from using the tools show that most participants preferred the Non-Contextual Summary, followed by the Contextual Summary, with the Metadata-based Snippet being the least preferred option (see Table~\ref{tab:auto_results}). The average score for ease of use is 4.4, while the average score for usefulness is 3.4. This indicates that evaluators found our tools easy to use but remained neutral regarding their overall usefulness.

\section{Related Work}
Automatic summarization in software engineering (SE) has attracted significant attention in recent years. Early approaches focused on generating human-readable commit messages or comments for code using sequence-to-sequence models and extractive methods like CodeSum, AST-based summarization, and attention-augmented transformers \cite{webis_abstractive_snippet, strzelecki2019featured, shangwenwang2023interpretation}. These techniques laid the groundwork for summarizing code and supporting developer comprehension tasks. Researchers also tackled the problem of PR summarization, framing PR description generation as a text summarization task. Notable work includes a T5-based model trained on over 33,000 PRs, which was shown to significantly outperform baseline systems using metrics such as ROUGE and BLEU \cite{sakib2024automaticpullrequestdescription,strzelecki2019featured}. These studies underscore the potential of LLMs for enhancing collaborative development via automated PR summarization. However, little work has explored the summarization of links embedded in PRs, particularly with respect to preserving link-level context in automated tools. Our approach diverges from prior studies by explicitly integrating link previews into the summarization pipeline-providing context-sensitive summaries that directly address external resources referenced in PRs.

\section{CONCLUSION AND FUTURE WORK}
We present \textsc{AILinkPreviewer}, an automatic link preview tool that leverages pull request context. In our evaluation, it outperforms LLMs without context or metadata-based summaries. While evaluators found it easy to use and relatively useful, they preferred the context-free LLM overall. In the future, we will improve the prompt for our summary as well as enable navigation of the relevant section of the webpage based on the context.

\section*{Acknowledgment}
We gratefully acknowledge the financial support of: (1) JSPS for the KAKENHI grants (JP25K22845); (2) Japan Science and Technology Agency (JST) as part of Adopting Sustainable Partnerships for Innovative Research Ecosystem (ASPIRE), Grant Number JPMJAP2415, and (3) the Inamori Research Institute for Science for supporting Yasutaka Kamei via the InaRIS Fellowship.

\bibliographystyle{IEEEtran}
\bibliography{references}

\end{document}